\documentstyle[preprint,prb,aps]{revtex}  
\begin{document}                

\title{Properties of High-Tc Single Crystals as Natural Interferometers 
in the $THz$ Frequency Range}
\author{  G. Hollauer$^a$ and Mauro M. Doria$^b$ }
\address{
$^a$ Departamento de F\'{\i}sica, Pontif\'{\i}cia Universidade 
Cat\'olica do Rio de Janeiro, Rio de Janeiro 22452-970 RJ, Brazil.\\
$^b$ Instituto de F\'{\i}sica, Universidade Federal Fluminense,
Campus da Praia Vermelha, Niter\'oi 24210-340 RJ, Brazil.}
\maketitle

\begin{abstract}

We consider oblique incidence of  (p)TM-polarized wave   on 
the anisotropic superconducting slab, immersed on a dielectric
media, such that its uniaxial (c) axis is perpendicular to
the surfaces.
The below and above plasma frequency transmissivity patterns are studied
and several of its properties determined, 
within the context of the Maxwell-London theory.
Below, the regime is  attenuated for any incident angle, and
there is a transmissivity maximum, quite pronounced 
in case of a very high external dielectric constant.
Above, a propagative regime exists where
the superconductor is a natural optical resonator, and we show 
here that the  minimum of the transmission peaks are
modulated by an envelope function associated to the Brewster 
condition.
We propose this set-up to obtain light with an
extremely small transverse wavelength inside
the superconductor.
\end{abstract}

\section{Introduction}

Conductors, well understood by the free-electron picture,
undergo a transition from opacity to transparency at the plasma
frequency. 
The effect, named after  the frequency range where it occurs in metals,
is known by ultraviolet transparency\cite{BOHREN}.
Much below the plasma frequency the properties of conductors are dominated
by dissipation resulting into an exponentially damped  electromagnetic fields 
away from  the surface,  and so,
nearly no light  penetration into the bulk (skin effect).
Similarly, light propagation in superconductors is also restricted to 
a strip near the surface, although by somewhat different reasons.
Ideal superconductors do not  dissipate, and
the exclusion of electromagnetic fields from 
the bulk is essentially a consequence of the Meissner effect
that forbids a magnetic field in its interior.
Thus light propagation in superconductors is also confined to a strip
below the surface, characterized by the
London penetration depth.
Indeed measurements involving the interaction of light with  superconductors,
below the pair-braking threshold, are surface effects, 
like the surface impedance\cite{GITTLEMAN} and 
the thin film plasmon\cite{BUISSON,DUNMORE}.
This is all in  agreement with the view of no plasma frequency below
the gap, thus condemning the superconductor to opacity throughout  the 
relevant frequency range, that is, much below the frequency associated to
the  gap\cite{MARTIN}.
Recently this view has completely changed due to the
infrared reflectivity measurements\cite{KOCH} of
c-axis polarized light done in 
several high-Tc superconductors, such as
$La_{2-x}Sr_xCuO_4$\cite{TAMASAKU,GERRITS,KIM}, 
$YBa_2Cu_3O_{8-x}$\cite{HOMES}, 
and $Bi_2Sr_2CaCu_2O_8$\cite{TAJIMA,OPHELIA1,OPHELIA2,MATSUDA}.
These measurements clearly indicate transparency 
starting  at a critical plasma frequency, 
hereafter called  $\omega_p$.
The startling novelty  is the evidence of a  plasma mode
below the energy gap, 
a  collective  charge oscillation 
between  the $CuO_2$ planes, that was not considered in
the early days arguments\cite{ANDERSON} when
such modes were proven  forbidden to isotropic superconductors.
Similarly to conductors, the reflectivity of  layered superconductors 
changes  dramatically depending whether
the frequency range studied is below or above $\omega_p$.
While below $\omega_p$ the behavior is the standard
one, namely amplitudes decay exponentially from
the surface\cite{DPB}, this is not the case above, where
surprisingly, the anisotropic (ideal) superconductor
behaves similarly to an anisotropic dispersive dielectric
medium.
The recent theoretical literature\cite{ARTEMENKO,TACHIKI,POKROVSKY} 
has discussed  several  unexpected properties of the
high-Tc ceramic single crystals, such as fiber-optic behavior,
namely light is totally evanescent outside the slab, and
in its interior, propagation can be pictured
through ray optics with multiple internal reflection at the 
interfaces\cite{DHPB}.

In this paper we consider a single crystal  
with the $CuO_2$ planes parallel to the external interface where a c-axis 
polarized $THz$  source sheds light at an oblique angle of incidence.
Single crystals are easily grown in this geometry that
we claim to work as a natural resonator in
the $THz$ frequency region\cite{ARTEMENKO,TACHIKI}.
Here we obtain  several of its  transmissivity characteristics 
below and above $\omega_p$ for this
particular geometry.
The present work applies at least for extremely low temperatures, 
where losses are negligible.

The study of  the reflectivity at oblique incident on a 
superconducting slab has been previously considered
by Artemenko and  Kobel'kov\cite{ARTEMENKO2}
in the context of kinetic equations for Green functions generalized
to the case of layered superconductors with weak interlayer
coupling.
The present paper is a more detailed study of such system,
where several new features are pointed out.
In particular  our study of light interaction with the anisotropic 
ideal superconductor is done in  the context of the  London-Maxwell
theory, the simplest possible framework to
understand the basic features of such systems
near the plasma frequency.
M. Tachiki et al.\cite{TACHIKI},
have  also obtained the
light reflectivity and transmissivity through a film,
however for a geometry distinct  from ours.
Notice that here the dielectric-superconductor interface has its normal  
along the c-axis, whereas in their case, it is
orthogonal to the c-axis.

Recently a  Fabry-Perot resonator was 
constructed with $YBCO$ films\cite{PORTEANU,FEENSTRA} and used to 
determine some of the superconductor properties, such as the complex 
conductivity, in the $GHz$ region.
When an integer number of half-wavelengths matches 
the distance between the two $YBCO$ films, 
light is transmitted though this interferometer.
This distance is an
adjustable parameter that allows for the search of
resonances at a given frequency.
The simplest interferometer of all 
is the single slab, in our case a high-Tc single crystal immersed 
on a dielectric medium.
The applications of the single slab interferometer are limited in 
comparison to the Fabry-Perot resonator because the thickness cannot be 
adjusted at will, although varying
the angle of incidence offers some degree of choice  in this sense.

We choose a coordinate system (see fig[\ref{fig1}]) where the interfaces
between the high-Tc single crystal  and  the dielectric media are at 
$x=d/2$ and $x=-d/2$.
According to our choice of crystal geometry the c-axis is 
just the x-axis.
The plane $x-z$ corresponds to  the so-called {\it plane of incidence}.
Thus fields are plane waves given by
$\exp{[-i(q_x x + q_z z - \omega t)]}$,  and
$\exp{[-i(\tilde q_x x + \tilde q_z z - \omega t)]}$, 
in the superconductor and in the dieletric medium, respectively.
Below we list  the remaining parameters used in the present model:
the angle of incidence $\theta$,
the vacuum wave number, defined as $k \equiv \omega /c$,
the two London penetration lengths,
transverse ($\lambda_{\perp}$), and longitudinal ($\lambda_{\parallel}$)
to the surfaces; the dielectric constant of the non-conducting media
exterior to the superconductor, $\tilde \varepsilon$; the film thickness $d$; 
and the distance between two consecutive $CuO_2$ planes $a$;
the  frequency independent dielectric constant
$\varepsilon_s$, the simplest phenomenological choice for the
superconductor's static dieletric constant along the c-axis.
This constant takes into account the contribution from high-energy 
interband processes and the atomic cores \cite{TANNER}

This paper is organized as follows.
In section \ref{section1},
we introduce the London-Maxwell theory, the Fresnel equation and derive
some noticeable properties below and above $\omega_p$.
In section \ref{section2}  we analyze the transmission patterns
expected for the two kinds of regimes characterized by 
$q_x^2$, the transverse wavenumber inside the superconducting slab: 
the attenuated ($q_x^2 < 0$) and the propagative ($q_x^2 > 0$).
Based on the the present model we  discuss on \ref{section3}
the optical properties  of the high-Tc ceramic  
$La_{2-x}Sr_xCuO_4$, $x=0.16$, at extremely low temperatures.
Finally in section \ref{section4} we summarize our major 
results.

\section{Basic Theory Above and Below $\omega_p$}\label{section1}

The simplest possible theoretical
framework able to describe  light propagation on the anisotropic 
superconductor, below and above $\omega_p$, is 
the London-Maxwell theory.
The approach is limited to
energies much lower than the pair breaking threshold.
The dynamics of the electromagnetic fields is described  
by the Maxwell's equations,
\begin{eqnarray}
{\bf \nabla} \cdot \;{\bf E} &=& {q\over{\epsilon_0}}\; (n_s({\bf x},t)-n_{0s}) \\
{\bf \nabla} \cdot\; {\bf H} &=& 0\\
{\bf \nabla} \times \;{\bf E} &=& -
\mu_0 {{\partial{\bf H}}\over{\partial t}}\\
{\bf \nabla} \times \; {\bf H} &=&  {\bf J} +
{{\partial{\bf D}}\over{\partial t}} 
\end{eqnarray}
In this paper we assume a low temperature regime,
and so, ignore any normal current contribution, thus
charge transport is just of the superfluid kind, described
by the first London equation,
adapted to the anisotropic problem.
\begin{eqnarray}
\mu_0 \lambda_{\parallel}^2 
{{ \partial {\bf J}_{\parallel} } \over
{\partial t}} \;=\; {\bf E}_{\parallel},
\quad 
\mu_0 \lambda_{\perp}^2 
{{\partial {\bf J}_{\perp}}\over{\partial t}} =  {\bf E}_{\perp}
\end{eqnarray}
where ${\bf E}_{\parallel}$, ${\bf J}_{\parallel}$ and
${\bf E}_{\perp}$, ${\bf J}_{\perp}$ 
are the field and supercurrent components
parallel and perpendicular to the interfaces,
respectively.

The time dependent electromagnetic coupling of the charged superfluid 
induces density fluctuations, a new feature not present
in the original formulation of London theory.
Notice that in the above London equation the supercharge density  is  
uniform and constant in space, $n_{0s}$ since 
the London penetration depths are
$\lambda_{\parallel}^2 = m_{\parallel}/n_{0s} q^2 \mu_0$
and 
$\lambda_{\perp}^2 = m_{\perp}/n_{0s} q^2 \mu_0$,
However coupling through Maxwell's equations forces
the supercharge density, $n_s({\bf x},t)$,
to fluctuate in time and space while  only the
neutralizing positive background remains uniform
and constant: $n_{0s}$.
This is  a linear  theory that simply ignores second order position and 
time dependent corrections to the density.
Such corrections are to a  time independent process,
which always has  uniform and constant density,
$( n_s({\bf x},t) = n_{ 0s} )$. 
For the case of interest here, namely wave propagation,
this neutrality is not possible, and, according to Gauss' law, 
the supercharge density  is found no longer constant and 
uniform.

At this point we introduce
the time dependence $\exp({i\;\omega\;t})$ to all fields,
assuming that the light frequency is much smaller than the
frequency associated to the superconducting  gap.
Introducing the first London equation,
\begin{eqnarray}
i\;\omega\;\mu_0 \lambda_{\parallel}^2 \,
{\bf J}_{\parallel} = {\bf E}_{\parallel}, \qquad
i\;\omega\;\mu_0 \lambda_{\perp}^2 \,
{\bf J}_{\perp} =  {\bf E}_{\perp},
\end{eqnarray}
into  Maxwell's equations, gives that,
\begin{eqnarray}
{\bf \nabla} \cdot \;{\bf D} &=& 0 \\
{\bf \nabla} \cdot\; {\bf H} &=& 0\\
{\bf \nabla} \times \;{\bf E} &=& - i\omega \mu_0 {\bf H}\\
{\bf \nabla} \times \; {\bf H} &=&  i \omega {\bf D} 
\quad \mbox{where} \quad
{\bf D} = \varepsilon_{s} \epsilon_0{\bf E} -i{\bf J}/\omega.
\end{eqnarray}
where the continuity equation,
$ {\bf \nabla} \cdot \;{\bf J} + i \omega q\; n_s = 0 $
has been used.
The  superconductor's  dielectric constant is
tensorial, ${\bf D} =  \epsilon_{s} {\bf E} -i{\bf J}/\omega =
\epsilon_0 {\bf \varepsilon} \cdot {\bf E} $. 
\begin{eqnarray} 
{\bf \varepsilon}  =  
\pmatrix{ \varepsilon_{\perp}  & 0 & 0  \cr 0 & \varepsilon_{\parallel}& 0  
\cr 0 & 0 & \varepsilon_{\parallel}  \cr } \quad
\varepsilon_{\perp}= \varepsilon_{s}- {1 \over {(k\lambda_{\perp})^2}} \quad
\varepsilon_{\parallel}=\varepsilon'_{s}-{1\over {(k\lambda_{\parallel})^2}}, 
\label{tensor}
\end{eqnarray}
The incident light wavelength, $2\pi/k$, is assumed much larger 
than the London penetration depth along the surface,
thus rendering an always negative dielectric  constant component
$\varepsilon_{\parallel}$.
\begin{eqnarray} 
\varepsilon_{\parallel} \approx -{1\over {(k\lambda_{\parallel})^2}},
\quad k\lambda_{\parallel} \ll 1
\label{epsp}
\end{eqnarray} 
A consequence of an extremely high anisotropy is that a similar property
does not hold perpendicularly to the film. 
Consequently the  incident light wavelength  matches the transverse London 
penetration depth even much below the pair breaking threshold.
Thus  the present model's  description near  
$\omega_p$  has physical content
and we use it to describe
phenomena  at the frequency window 
where  this dielectric component $\varepsilon_{\perp}$  flips  sign.
\begin{eqnarray}
\varepsilon_{\perp}(\omega=\omega_p)=0 \qquad \omega_p =
{{c}\over {\sqrt{\varepsilon_s}\lambda_{\perp}}}  \label{plo}
\end{eqnarray}

The interesting polarization is the so-called $p$ or $TM$  
(Transverse Magnetic), the only one that  probes properties
of the  dielectric component $\varepsilon_{\perp}$.
We notice, for this polarization,  a superficial charge density 
build-up at the interfaces whose oscillation yields interesting low frequency 
properties\cite{DHPB}.
Its field components $(E_x, H_y, E_z)$ imply on  a transverse current 
$J_x$, not present in the external insulating dielectric media,
that lead to this superficial charge density.
The $s$ or $TE$ (Transverse Electric) polarized wave, with its field 
components $(H_x, E_y, H_z)$, only probes $\varepsilon_{\parallel}$, 
and consequently fields are confined to a strip near the interfaces.
for this reason it is not  considered in this paper.

Long before the advent of Maxwell's equations, Fresnel had 
developed a geometrical method to explain
the properties of light propagation
on anisotropic dielectric media characterized by more than
one refraction index. 
It is straightforward to derive from the present Maxwell-London theory 
the so-called Fresnel's equation of wave normals,
\begin{eqnarray}
{{{q_z}^2 }\over{\varepsilon_{\perp    }} } + 
{{{q_x}^2 }\over{\varepsilon_{\parallel}} } = k^2 
\label{eqs1}
\end{eqnarray}
inside the superconductor ($-d/2<x<d/2$), and,
\begin{eqnarray}
{\tilde q_z}^2  +  {\tilde q_x}^2  = \tilde\varepsilon k^2 
\label{eqs2}
\end{eqnarray}
in  the dielectric media ($x<-d/2$; $x>d/2$).
According to   Fig.(\ref{fig1}) we solve Eq.(\ref{eqs2}) for
the incident wave in the  dielectric,
using  the  previously introduced parameters:
$\tilde q_x = - k \sqrt{\tilde \varepsilon}\cos{\theta} $, 
and  $\tilde q_z = k \sqrt{\tilde \varepsilon}\sin{\theta}$.
Snell's law states that the wavenumber parallel to the surface is 
the same for the incident, reflected and transmitted waves:
$ q_z = \tilde q_z $.
Finally we obtain  an expression, central  for our purposes here,
that gives  $q_x$, the transverse wavenumber inside the superconductor,
in the approximation described by Eq.(\ref{epsp}).
\begin{eqnarray}
{q_x}^2= {1 \over {{\lambda_{\parallel}}^2}}
\big( {{\tilde \varepsilon}\over{\varepsilon_{\perp}(\omega)}} \sin^2{\theta} -1 \big) \label{qx2}
\end{eqnarray}
This equation is a valuable tool 
to understand many of the unusual properties of light interaction with
the (ideal) anisotropic superconductor.
For instance, the normal incidence ($\theta=0$) is  always attenuated
exponentially  decaying over a distance $\lambda_{\parallel}$.
Next we derive  further results, in special several
frequency parameters useful to understand the
properties of wave propagation inside the superconducting
slab.

\subsection{ Below $\omega_p$ }\label{bo}
 
For $\varepsilon_{\perp}<0$ the transverse wavenumber inside
the superconductor, $q_x$, given by Eq.(\ref{qx2}),
is imaginary, and so, the TM
wave is attenuated in this frequency range. 
The attenuation length shortens as the frequency
increases from zero to $\omega_p$.
To see this, notice that deeply below $\omega_p$, the 
transverse dielectric constant is approximatly given by 
$\varepsilon_{\perp} \approx -1 / (k\lambda_{\perp})^2$, 
similarly to Eq.(\ref{epsp}), resulting that,
${q_x}^2 \approx -[ (k\lambda_{\perp})^2 \sin^2{\theta} + 1]/
{\lambda_{\parallel}}^2 $.
Thus the attenuation length, tipically 
of order $\lambda_{\parallel}$ for  low frequencies, 
diminishes as the frequency increases,
and eventually vanishes at $\omega_p$.

\subsection{ Above $\omega_p$ }\label{ao}
The regime is propagative for oblique incidence,
no matter how small $\theta$ is, provided that
a frequency window is selected sufficiently
close to the plasma frequency ($\varepsilon_{\perp} \sim 0^{+}$).
Thus a frequency window exists,  starting at $\omega_p$ and ending at 
$q_x=0$, this last condition associated to the frequency below, 
obtained from  Eq.(\ref{qx2}). 
\begin{eqnarray}
\omega_0 = 
{{\omega_p}\over{\sqrt{1-{{\tilde \varepsilon }\over{\varepsilon_s}}
\sin^2{\theta}}}} \label{om0}
\end{eqnarray}
The propagative frequency window,
$(\omega_p, \omega_0(\theta))$ exists until a critical angle is reached,
thus defined only for an angular window $(0, \theta_c)$.
\begin{eqnarray}
\theta_c = \left \{ \begin{array}{ll}
\arcsin{\sqrt{\varepsilon_s/\tilde \varepsilon}} & 
\quad \tilde \varepsilon > \varepsilon_s \\
\pi/2 & \quad \tilde \varepsilon < \varepsilon_s \end{array} \right.
\end{eqnarray}
Notice that for  $\tilde \varepsilon > \varepsilon_s $ 
there is another attenuated region above the critical angle,  
$\theta_c < \theta \le \pi/2$.

A well known property of oblique TM wave, incident on
the interface between two dielectric, is of no reflection
at the so-called Brewster angle.
We show here that for the interface dielectric-superconductor
this  property of no reflection occurs in 
the whole propagative angular region  $(0,\theta_c)$.
This is a dispersive medium property,
that for any angle of incidence allows for the
choice of a refraction index by tuning the frequency, 
such that the Brewster condition is met.
This Brewster condition of no reflection is given below,
following standard arguments, obtained  from the study of a plane wave 
incident on the interface between the two media.
\begin{eqnarray}
F \equiv {{\tilde \varepsilon}\over{\tilde q_x}} 
{{q_x}\over{\varepsilon_{\parallel}}}, \quad F = 1
\label{F}
\end{eqnarray}
Before proceeding any further,
we briefly  show that
the above relation yields  the Brewster angle in
case of a single interface between two non-conducting dielectric media.
For the sake of the argument  ignore 
the frequency dependence of $\varepsilon_{\parallel}$ and solve Eq.(\ref{F}): 
$\cos{\theta}/\sqrt{\varepsilon_{\parallel}}
= \cos{\tilde \theta}/\sqrt{\tilde \varepsilon}$,
where $\tilde \theta$ and $\theta$ are the incident
and transmitted angles, respectively.
Added of Snell's law,  
$\sqrt{\varepsilon_{\parallel}} \sin{\theta} 
= \sqrt{\tilde \varepsilon} \sin{\tilde \theta}$,
one finally obtains the Brewster angle condition,
$\theta + \tilde \theta = \pi/2$.
To obtain this result we have introduced into Eq.(\ref{F}) that
$ {q}_x = - k \sqrt{\varepsilon_{\parallel}} \cos{\theta}$ and 
$ {\tilde q}_x = - k \sqrt{\tilde \varepsilon} 
\cos{\tilde \theta}$.
Back to the superconductor-dielectric interface, 
square Eq.(\ref{F}) and introduce both
Eq.(\ref{qx2}) and Eq.(\ref{epsp}).
The Brewster condition is determined by 
the transverse wavenumber and the frequency 
obtained at a given $\theta$.
\begin{eqnarray}
q_{x,B}^2 &=& \lbrack 
\sqrt{(\alpha-1)^2 - 4 \alpha \beta } + (\alpha-1)
\rbrack/2\lambda_{\parallel}^2 \\
\omega_B^2 &=& \omega_p^2 \lbrack 
\sqrt{(\alpha-1)^2 + 4 \alpha \beta } - (\alpha-1)
\rbrack/ (2\beta) \\
\alpha &=& {{\varepsilon_s}\over{\tilde \varepsilon}}
\big( {{\lambda_{\perp}}\over{\lambda_{\parallel}}}\big )^2 \cos^2{\theta} \\
\beta &=& 1 - {{\tilde \varepsilon}\over{\varepsilon_s}} \sin^2{\theta}
\end{eqnarray}
We find that  properties of the Brewster condition 
depend whether the ratio $\tilde \varepsilon/\varepsilon_s$
is smaller or larger than one.
Let us restrict the discussion, in both cases, to 
the highly anisotropic
superconductors, defined through the condition 
$\sqrt{\varepsilon_s /\tilde \varepsilon}
(\lambda_{\perp}/ \lambda_{\parallel}) \gg 1$.
For $\tilde \varepsilon/\varepsilon_s < 1$ 
the Brewster frequency varies from $\omega_B \approx \omega_p$, for 
$\theta=0^{+}$, to $\omega_B \approx \omega_0$,  for $\theta=\pi/2$.
The transverse wavenumber associated to each of these extreme angles is 
$ q_{x,B}(0) \approx \sqrt{\varepsilon_s /\tilde \varepsilon}
(\lambda_{\perp}/ \lambda_{\parallel}^2)$,
and $q_{x,B}(\pi/2)\approx 0$, respectively.
For $\tilde \varepsilon/\varepsilon_s > 1$ the propagative
regime also begins at $\theta=0^{+}$, like the previous case,
and ends at $\theta_c$ with the Brewster frequency
$\omega_B \approx \omega_p \sqrt{\alpha'/(\alpha'-1)}$,
$\alpha' \equiv (\varepsilon_s /\tilde \varepsilon) 
(1-\varepsilon_s /\tilde \varepsilon) (\lambda_{\perp}/ 
\lambda_{\parallel})^2 $.
The corresponding transverse wavenumber is  
$q_{x,B}(\theta_c) \approx \sqrt{\alpha'}/\lambda_{\parallel}$,

In this paper we are interested on  the 
finite thickness superconducting slab
immersed on a dielectric medium. 
A well-known optical property
of a film\cite{YEH} is the resonant condition, that leads
to total light transmission through the film,
namely  an integer number of 
transverse half-wavelengths perfectly matching  the thickness,
$N\; \lambda_x/2 = d$.
This condition on Eq.(\ref{qx2}) gives that,
\begin{eqnarray} 
q_x = N{{\pi}\over{d}} \quad 
\omega_N = \omega_p \sqrt{\frac
{1+ \big(\frac{N\pi\lambda_{\parallel}}{d}\big)^2}
{(\frac{\omega_p}{\omega_0})^2 + 
\big(\frac{N\pi\lambda_{\parallel}}{d}\big)^2}}
\label{omN}
\end{eqnarray}
where $N$ is the number of half-wavelengths.
Notice that the propagative frequency window is swept
backwards for increasing $N$, i. e.,
the maximum propagative frequency of Eq.(\ref{om0}) is associated to the 
smallest integer, $N=0$.
Near the minimum propagative frequency, $\omega_p$, $N$ is
large, but thre is a maximum value 
associated to the limit of this theory's validity, which cannot describe
a transverse wavelength so small as the interlayer distance $a$.
Another model, able to describe individual layers, 
such as the Lawrence-Doniach model\cite{DONIACH}, must be considered
in this limit.
\begin{eqnarray} 
N_{max} = 2 d /a \quad 
{q_x}_{N_{max}} = {{2\pi}\over{a}} \quad 
\omega_{N_{max}} = \omega_p \sqrt{\frac
{1+ \big(\frac{2\pi\lambda_{\parallel}}{a}\big)^2}
{1 -\frac{\tilde\varepsilon}{\varepsilon_s}\sin^2{\theta} + 
\big(\frac{2\pi\lambda_{\parallel}}{a}\big)^2}}
\label{omN3}
\end{eqnarray}
In the usual cases  this maximum frequency falls extremely close to the 
plasma frequency,  $\omega_{N_{max}} \sim \omega_p^{+}$, a direct
consequence of a longitudinal London penetration depth
much larger than the inter-plane distance:
$\lambda_{\parallel} \gg a$.

It follows from Eq.(\ref{omN}) the enlargement of the  distance 
between two consecutive resonances, $\omega_{N-1}-\omega_N$, 
for an increasing  propagative window $(\omega_p, \omega_0)$.
This is a valuable remark, considering that 
the choice on  $\omega_0$ depends on other parameters such as an
appropriate angle of incidence, $\theta$ and (or) the external
dielectric medium constant, $\tilde \varepsilon$.
Let us check that indeed this is the case for two
special cases of Eq.(\ref{omN}):
{\it (i)} Near $\omega_p$, one has that  
$d/N_{max}\pi\lambda_{\parallel} \ll 1$ and, in this
limit, it holds that
$\omega_{N-1}-\omega_{N} \approx \omega_p \lbrack 
1- (\omega_p/\omega_0)^2 \rbrack
(d/\pi\lambda_{\parallel})^2/N^3 $.
{\it (i)} Near $\omega_0$ we just provide the distance 
between the square of the last two maxima,
$\omega_0^2-\omega_1^2 = \omega_0^2
(\pi\lambda_{\parallel}/d)^2 \lbrack 1 - (\omega_p/\omega_0)^2 \rbrack /
 \lbrack  (\omega_p/\omega_0)^2 + (\pi\lambda_{\parallel}/d)^2\rbrack $.
Both limits clearly indicate that for increasing $\omega_0$, 
and so small  ratio $\omega_p/\omega_0$, the distance between 
consecutive frequency maxima 
also increases.

\section{ Transmission Pattern Analysis} \label{section2}

Applying the standard arguments of  continuity of the transverse fields, 
$H_y$ and $E_z$, one obtains the ratio
between the bottom and the top normal components of the
Poynting's vector\cite{YEH}, and so,
the transmissivity coefficient, $T$,
that measures the transmission of energy through the film.
For the case of identical  top and bottom dielectric media, 
$T$ is a sole function of the  Fresnel reflection coefficient 
($\rho$) of a single surface, and of the phase parameter ($\eta$).
\begin{eqnarray}
{\rho ={{1-F}\over {1+F}}  
\quad \eta =\exp{(i\tilde q_x d)} \quad T =  \big| {{\eta(1-{\rho^2})}\over {1-{\eta^2}{\rho^2}}}\big|^2
};\label{eqs4}
\end{eqnarray}
where $F$ has been previously defined (Eq.(\ref{F})).

\subsection{The Attenuated Regions}

There are two  possible attenuated regions for a fixed $\theta$, 
below the plasma frequency,
$(0< \omega < \omega_p)$, and  above 
the propagative window, $(\omega > \omega_0)$.
Both regions display the common feature 
that $1-(\tilde \varepsilon/\varepsilon_{\perp})\sin^2{\theta} > 0$.
In these regions the transverse wavelenth, $q_x$, is imaginary
thus being interpreted as a penetration depth.
The ratio defined in Eq.(\ref{F}) is imaginary,
the phase parameter $\eta$ real, and so,
the transmissivity coefficient of Eq.(\ref{eqs4}) becomes,
\begin{eqnarray}
T = {{8 \; f^2}\over
{\big(1+f^2\big)^2 \cosh{(2 d/ l_x)}-
\big( 1-6 f^2 + f^4 \big)}}  \nonumber \\
f^2 = \big({{\omega\lambda_{\parallel}}\over{c}} \big)^2 
{{\tilde\varepsilon}\over{\cos^2{\theta}}}\big (
1 - {{\tilde \varepsilon}\over{\varepsilon_{\perp}}} \sin^2{\theta} \big) 
\quad l_x = {{\lambda_{\parallel}} \over
{\sqrt{1 - {{\tilde \varepsilon}\over{\varepsilon_{\perp}}} 
\sin^2{\theta} } }}
\label{eqs7}
\end{eqnarray}
The transmissivity has a peak below the plasma frequency since 
at extremely small frequencies it is an increasing function, 
growing proportionally to $\omega^2$,
and very near $\omega_p$ it becomes a decreasing function that 
eventually vanishes at $\omega_p$.
We  obtain here the asymptotic limit 
of Eq.(\ref{eqs7}), in this last regime that shows 
the exponential decay of the transmissivity
below and extremely close to $\omega_p$.
For $\theta \ne 0$ the transmissivity is approximately described by
$T \approx 16 f^{-2}\exp{(-2d/l_x)}$ that leads to,
\begin{eqnarray}
T = 32 \big( 
{{\varepsilon_s}\over{\tilde \varepsilon}} 
{{\lambda_{\perp}}\over{\lambda_{\parallel}}}
\big)^2 {{1} \over {\tan^2{\theta}}}
\big({{\omega_p}\over{\omega}}-1 \big)
\exp{-\big({{d}\over{\lambda_{\parallel}}}
\sqrt{{{2\tilde\varepsilon}\over{\varepsilon_s}}}
{{\sin{\theta}}\over{ \sqrt{ {{\omega_p}\over{\omega}}-1}}}
\big)} 
\label{eqs8}
\end{eqnarray}


\subsection{ The Propagative Region}
In the propagative window, $(0, \theta_c)$  and 
$(\omega_p, \omega_0)$, one has that 
$(\tilde \varepsilon/\varepsilon_{\perp})\sin^2{\theta}-1 > 0$ and 
the  transmissivity becomes,
\begin{eqnarray}
T = {{8 \; F^2}\over
{\big( 1+6 F^2 + F^4 \big)-\big(1-F^2\big)^2 \cos{(2 q_x d)}}} \label{eqs9}
\end{eqnarray}
where  $F^2 = (\omega\lambda_{\parallel}\tilde\varepsilon/c\cos^2{\theta} )^2 
[(\tilde \varepsilon/\varepsilon_{\perp}) \sin^2{\theta} - 1]$,  
according to Eq.(\ref{F}),
and $q_x$ is given by Eq.(\ref{qx2}).
The transmissivity  maximum  $(T=1)$ corresponds to  $\eta=1$
$(q_x = N\pi/d)$, 
the resonance condition previously described (Eq.(\ref{omN})).

The Brewster condition is well defined in the context of
the envelope curve that fits all the transmissivity minima.
The maximum of this envelope function is a 
convenient definition of the Brewster condition.
The reason for this is very simple, when the  transmissivity minima 
and  maxima are equal no wave is reflected 
and the Brwester condition is met.
The transmissivity  minima occur for $\cos{(2 q_x d)} = -1$.
Hence the minimum transmissivity is given by 
$T_{min} = 4 F_{min}^2/(1+F_{min}^2)^2$
where $F_{min}$ is obtained from Eq.(\ref{F}) under the minimum wavenumber 
condition, $q_x = (N+1/2)\pi/d$.
These points correspond to an integer number of 
transverse half-wavelengths,
added of an extra one forth of a wavelength, matching the thickness.

>From its turn  the envelope curve,
defined by the continuous function 
$T_{env} \equiv 4 F^2/(1+F^2)^2$ of parameter $q_x$,
has a maximum at $F^2=1$, 
the Brewster condition previously discussed in Eq.(\ref{F}).
Obviously $T_{env}$
fits all the transmissivity minima within
the propagative window.
Near the Brewster point  interferometric
properties of the finite superconduciting slab 
are no longer useful.


\section{ Discussion} \label{section3}

For our discussion  we choose the high-Tc ceramic 
$La_{2-x}Sr_xCuO_4$, which has a single superconducting layer per unit cell,
following  Tamasaku, Nakamura, and
Uchida\cite{TAMASAKU} data on high-quality single crystals
of this superconductor. 
For the composition $x= 0.16$ the critical 
temperature is $T_c = 34 K$.
For $T = 8 K$ these authors obtain from the reflectivity data,
and Kramers-Kronig analysis, the transverse dielectric constant,
$ \varepsilon_{\perp} = \varepsilon_s - {\omega'_{\perp}}^2/\omega^2 -
({\omega_{\perp}}^2-{\omega'_{\perp}}^2)/[\omega(\omega+i\gamma)]$, 
fitted by the following parameters:
$\varepsilon_s = 25$, $\omega'_{\perp} = 8.4\; 10^{12} \;rad/s $, 
$\omega_{\perp} = 9.0 \; 10^{12} \; rad/s $
and $\gamma = 5.0 \; 10^{10} \; rad/s $.
The two frequency dependent terms in the dielectric
constant are, respectively, the contribution of
the condensed carriers and the thermally excited
quasi-particles.
In this paper losses are being totally discarded,
thus we extrapolate the above data to
$\gamma = 0$, thus obtaining
the transverse London penetration 
$\lambda_{\perp} = c/\omega_{\perp} = 33 \; \mu m$, and
the plasma frequency  $\omega_p = \omega_{\perp}/
\sqrt{\varepsilon_s} = 1.8 \; 10^{12}\; rad/s$.
For the  zero-temperature  London penetration length along the 
$CuO_2$ planes we take that
$\lambda_{\parallel}= 0.2 \mu m$ \cite{BLATTER}, thus yielding
an effective anisotropy of 
$\lambda_{\perp}/\lambda_{\parallel} = 165$.
The distance between two consecutive $CuO_2$ planes in this compound
is $a=0.7 \;nm$ \cite{CYROT}.
For illustration purposes we choose the slab sufficiently thin in order
to deal just with fairly small number of frequency resonances in 
the propagative window:  $d = 0.1\; \mu m$ and, consequently, 
$N_{max} = 285$, according to Eq.(\ref{omN3}).

Fig.(\ref{fig1})  provides
a pictorial intuitive view of a p-polarized wave 
incident on a superconducting slab immersed
on a dielectric media of dielectric constant $\tilde
\varepsilon$.
The anisotropic superconductor has its c-axis
perpendicular to the interfaces, and is characterized
by two-dielectric constants, the longitudinal  ($\varepsilon_{\parallel}$), 
monotonous and always negative
in any frequency window, and the transverse
($\varepsilon_{\perp}$), that flips sign, creating the
below and above  plasma frequency regimes.

Fig.(\ref{fig2}) displays the  below
plasma transmissivity for $\theta=20^o$, $50^o$, and $80^o$.
In this frequency range the transmissivity has a peak,
whose height is quite pronounced  due to  the 
choice of a nonconducting bounding medium of very high dieletric constant, 
namely $SrTiO_3$ \cite{BUISSON}: 
$\tilde \varepsilon \approx 2.0\;10^{4}$.
In case the slab is immersed in  air ($\tilde \varepsilon = 1$), 
the transmissivity maxima are $2.1 \; 10^{-4}$,
$4.5 \; 10^{-4}$, and 
$6.0 \; 10^{-3}$, for the above angles of incidence, respectively, 
all occuring at frequencies extremely close to
the plasma frequency $(\omega \approx 0.99\omega_p)$. 
Therefore the peak still exists in case of a small dielectric constant,
although very small.

Fig.(\ref{fig3}) and  Fig.(\ref{fig4}) show the  above $\omega_p$
transmissivity for $\theta=20^o$, and $80^o$, respectively.
The many spikes, observed in
these transmissivity patterns, are the resonances 
labeled by ther number of transverse half-wavelengths
that fit transversally into the slab.
In order to best visualize the spikes we have chosen the frequency window
to $(\omega_p, \omega(10))$.
Indeed the spikes are less dense near $\omega_0$ than near $\omega_p$ as 
shown in these two Figures.
Properties of the spikes, such as the density, are studied through
the functions,
\begin{eqnarray}
Q_1(N) \equiv {{\omega(N)}\over {\omega(N-1/2)-\omega(N+1/2)}} \\
Q_2(N) \equiv {{2}\over{T(N-1/2)+T(N+1/2)}}
\end{eqnarray} 
The first function provides information on the spike density.
In fact near $\omega_p$, where this density is high 
$d/\pi\lambda_{\parallel}N \ll 1$, one obtains that
$Q_1(N) \approx N^3 (\pi \lambda_{\parallel}/d)^2 / 
[ 1 - (\omega_p/omega_0)^2 ]$.
Near $\omega_0$ consecutive peaks are well separated and
$Q_1(N)$  is small.
The second function is just the ratio of the transmissivity maximum
and the average transmissivity taken between the two neighbor transmissivity 
minima.
In case the minima are not ptonounced, namely are approximatly one, 
$Q_2(N)$ becomes approximately equal to one.
For interferometric purposes the best
resolution for a spike is sought and , for this purpose, both
$Q_1(N)$ and  $Q_2(N)$  must be much larger than one 
in the  frequency range under investigation.
Fig.(\ref{fig5}) and  Fig.(\ref{fig6}) display the above
functions versus $\omega-\omega_p$,  for $\theta=20^o$, 
and $80^o$.
The frequency range displayed ranges from 
$\omega(N_{max}-1)$, very near
$\omega_p$, to  $\omega(1)$, the last spike before the end
of the propagative regime at
$\omega(0)$.
Fig.(\ref{fig5}) shows that $Q_1$ drops over many orders of magnitude for 
increasing frequency, confirming
that the  density of spikes is maximum  near  $\omega_p$
and mimimum  near $\omega_0$.
The minimum of  $Q_2$, shown in Fig.(\ref{fig6}) for
both angles,
has in tis mimimum the Brewster condition, 
since the transmissivity maximum and minimum  are equal to one.
Notice that  the frequency associated to the minimum of
$Q_2$  is just the maximum of the 
envelope function shown in Fig.(\ref{fig3}) and  Fig.(\ref{fig4}).
This frequency, $\omega_{B}$,
shows, in a explicit way, that the Brewster condition 
of no $TM$ reflected polarized
light is more than  just an angle, being  a set of parameters, 
including $\omega_{B}$ and its associated
angle of incidence.

Lastly Fig.(\ref{fig7}) displays $\omega_{B}-\omega_p$ 
vs. $\theta$.
For very small angle of incidence the Brewster frequency is
close to $\omega_p$ and for a grazing angle moves to an intermediate 
position in the propagative window.
This figure also shows the behavior of the lower
($\omega(N_{max})-\omega_p$) and upper 
($\omega_0-\omega_p$) limits of the propagative
window vs. $\theta$.

\section{Conclusion}\label{section4}

In this paper  we have studied  transmissivity of the
TM-polarized light through a superconducting slab
immersed on a dieletric media.
The superconductor is anisotropic and
the surface, where light is obliquely incident, is
orthogonal to the c-axis.
The present study is done in the context of the London-Maxwell theory,
which gives a good account of the
physical situation for scales larger than the interplane
separation.
The anisotropic superconductor has a tensorial dielectric
constant.
Along the surfaces the (longitudinal) component is negative because
the incident  beam wavelength is always much larger than the 
longitudinal London penetration depth.
However  the situation is quite distinct 
along the transverse direction.
The incident wavelength can be comparable to the transverse London 
penetration depth implying that  the
physical range where this dielectric constant component vanishes is an 
interesting one even though
we are only interested in phenomema taking place much
below the pair breaking threshold.

The effects of a plasma frequency along the c-axis
on the transmissivity patterns is studied here.  
Below the plasma frequency the regime is attenuated
and an interesting feature arises in case the slab
thickness is smaller than the longitudinal London penetration
depth.
The interaction between the two surfaces renders
a transmissivity  maximum that 
can be quite pronounced in case the external 
dielectric constant is very large.
In particular we have shown  here that the transmissivity vanishes 
exponentially below and very near to the plasma frequency.
Above the plasma frequency there is a frequency and angular
window where the regime is propagative inside the superconductor 
independently on how thick slab is.
Within this window, the superconducting slab works as a natural optical 
resonator displaying transmissivity resonances
labeled by the number of  transverse half-wavelength that exactly match the 
slab. 
A remarkable property of this regime is the transverse wavelength inside
the superconductor, which can be as small as 
desired up the order of the inter-plane separation.

TM-polarized light, obliquely incident on a surface between two dielectric 
media, has the property of  
no reflection at the so-called Brewster angle.
Similarly  for the dielectric-superconductor interface
a Brewster condition exists.
We find here that this Brwester condition of
total transmission exists
for any angle of incidence within the propagative range,
once the incident frequency is chosen suitably.
In summary we find that the transmissivity minima
are fitted by an envelope curve, whose maximum gives
the Brester condition.

The present work does not take into account losses
thus being restricted to extremely low superconductors 
where thermally activated normal carriers can be ignored.
In this contex we have used parameters of the high-Tc ceramic 
$La_{2-x}Sr_xCuO_4$ to show the transmissivity features
discussed here.

In  summary we have determined in this paper 
several of the transmissivity properties of an
anisotropic superconducting slab immersed on a
dielectric medium which has the startling property
of behaving, above the c-axis plasma frequency,
as an anisotropic dielectric for incident TM polarized
light.


\newpage
	     
\baselineskip = 2\baselineskip  
\begin{figure}
\caption{ A pictorial view of a TM polarized wave incident 
on a superconducting slab
immersed on a nonconducting dielectric medium. 
The multiple reflections occuring at the interfaces suggest
that an above $\omega_p$ situation is being represented here.}
\label{fig1}
\end{figure}
\begin{figure}
\caption{The below $\omega_p$ transmissivity  
of the $La_{2-x}Sr_xCuO_4$ superconductor as described in the text, 
is  shown here for three angles of incidence.
The quite pronounced transmissivity peaks should be credited 
to an exterior dielectric medium of elevated
constant, $SrTiO_3$: $\tilde \varepsilon \approx 2.0\;10^{4}\; rad/s$.}
\label{fig2}
\end{figure}
\begin{figure}
\caption{ The transmissivity is shown here in the window
$(\omega_p, \omega(10))$  for $\theta = 20^o$.
The reimaining parameters are those of the $La_{2-x}Sr_xCuO_4$ superconductor, 
as described in the text.
Notice that each spike is a resonance, labeled by 
the  number of transverse half-wavelengths that
exactly fit into the slab.}
\label{fig3}
\end{figure}
\begin{figure}
\caption{ The transmissivity is shown here in the window
$(\omega_p, \omega(10))$  for $\theta = 80^o$.
The remaining parameters are those of the $La_{2-x}Sr_xCuO_4$ superconductor, 
as described in the text.
Notice that each spike is a resonance, labeled by 
the  number of transverse half-wavelengths that
exactly fit into the slab.}
\label{fig4}
\end{figure}
\begin{figure}
\caption{ The ratio between the resonance frequency
and the distance between the two neighbor minima
is shown here for two angles of incidence in the
frequency window $(\omega(N_{max}-1), \omega(1))$.
The remaining parameters are those of the $La_{2-x}Sr_xCuO_4$ 
superconductor, as described in the text.
The frequencies $\omega_1$, $\omega_B$ and $\omega_{N_{max}-1}$ 
are indicated for the $\theta =20^o$ curve.}
\label{fig5}
\end{figure}
\begin{figure}
\caption{ The ratio between the maximum transmissivity (one)
and the the average transmissivity of the two neighbor minima
is shown here for two angles of incidence in the
frequency window $(\omega(N_{max}-1), \omega(1))$.
The remaining parameters are those of the $La_{2-x}Sr_xCuO_4$ 
superconductor, as described in the text.
The frequencies $\omega_1$, $\omega_B$ and $\omega_{N_{max}-1}$ 
are indicated for the $\theta =20^o$ curve.}
\label{fig6}
\end{figure}

\begin{figure}
\caption{ This figure shows the  lower
($\omega(N_max)-\omega_p$) and the upper 
($\omega_0-\omega_p$) limits of the propagative
window, as well as the Brewster frequency ($\omega_B-\omega_p$),  
as a function of the angle of incidence ($\theta$).}
\label{fig7}
\end{figure}

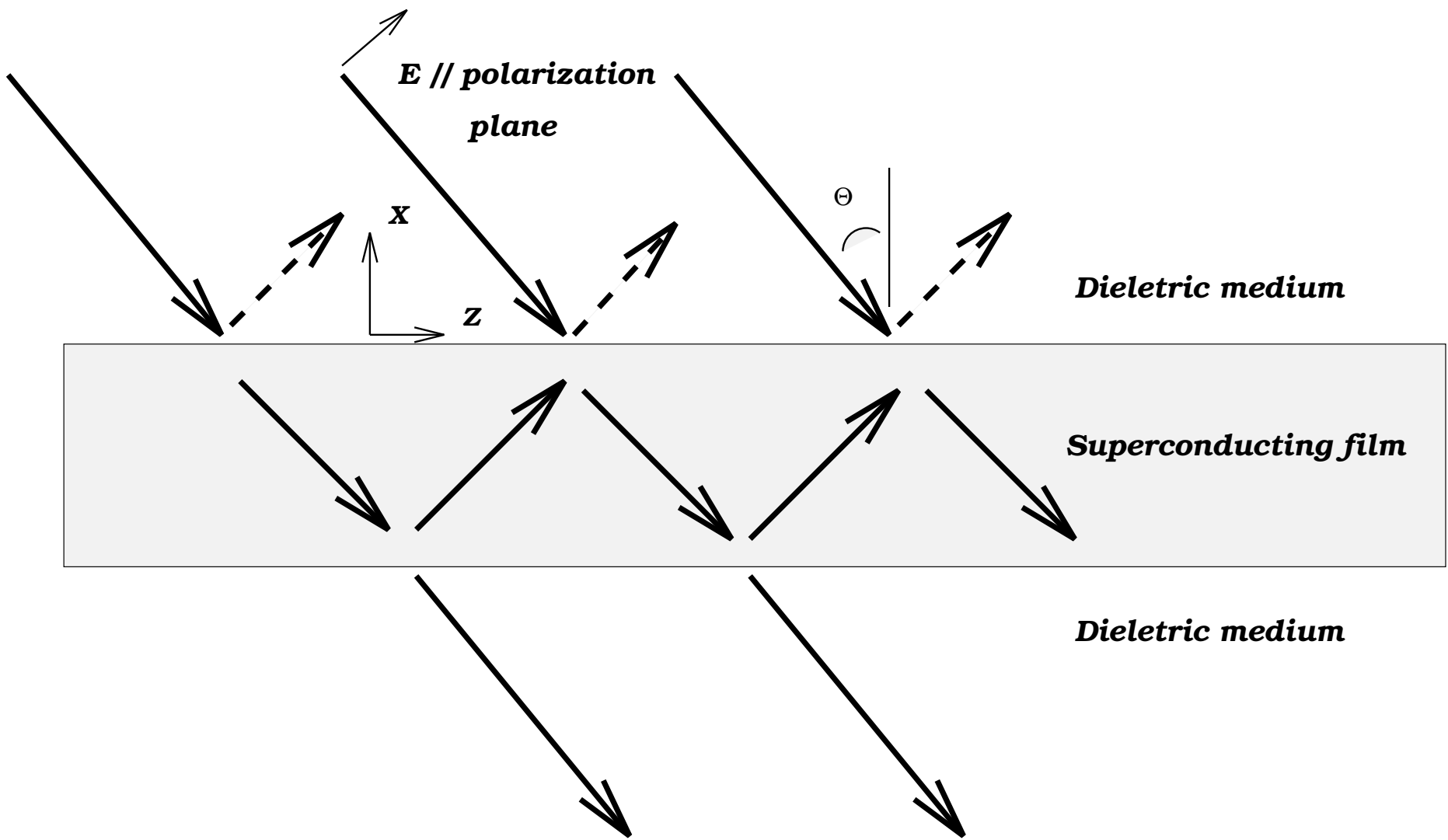
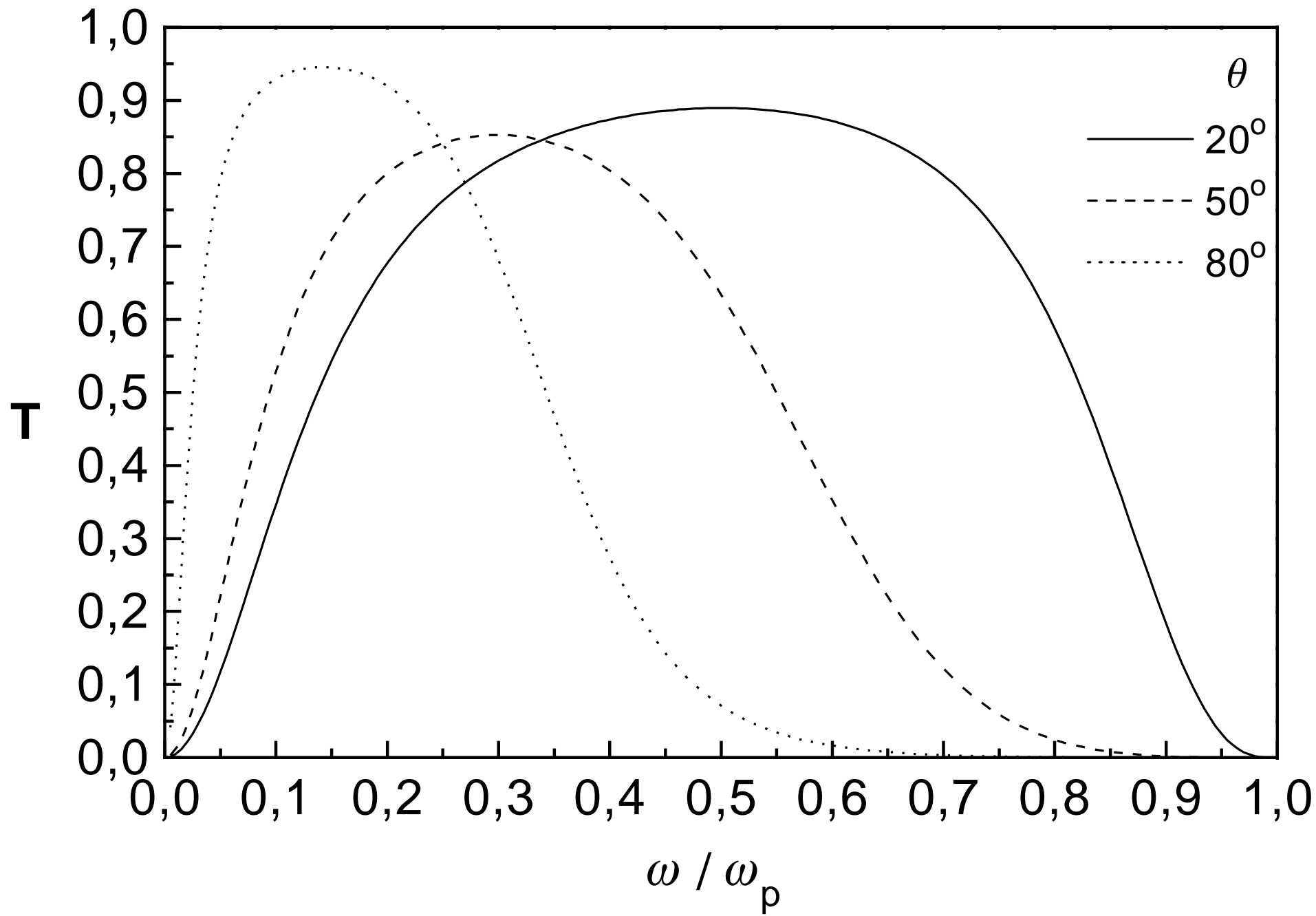
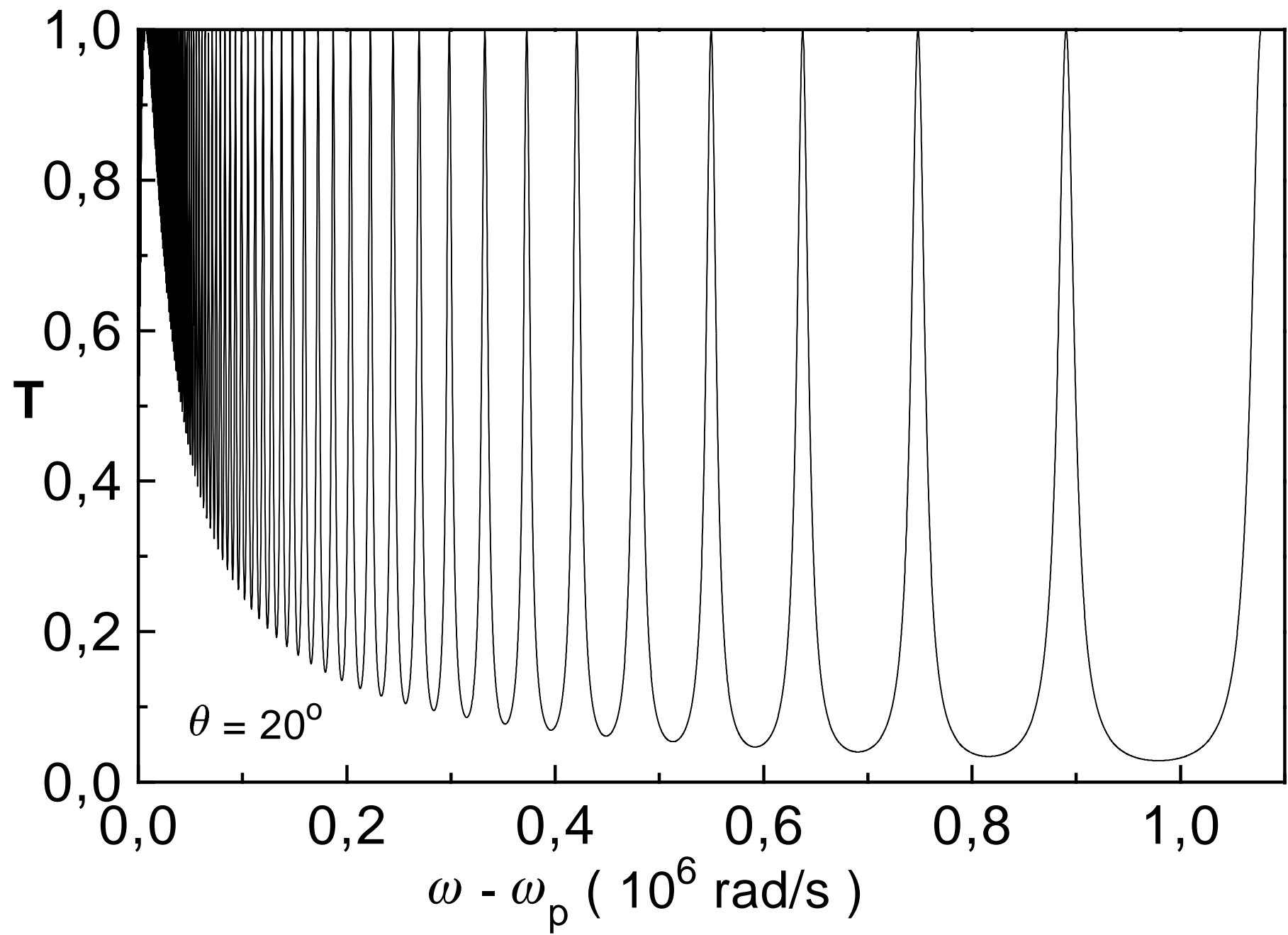
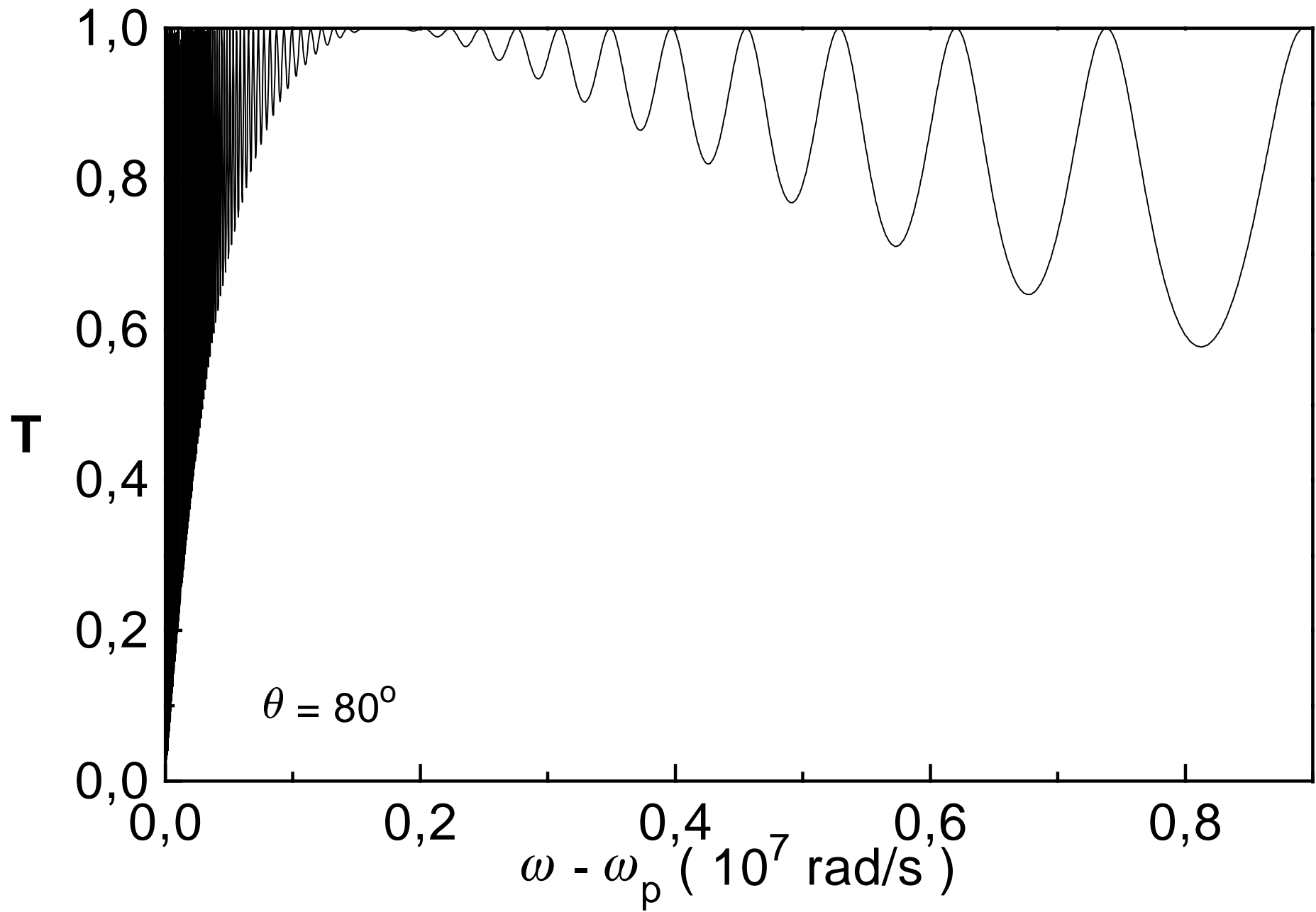
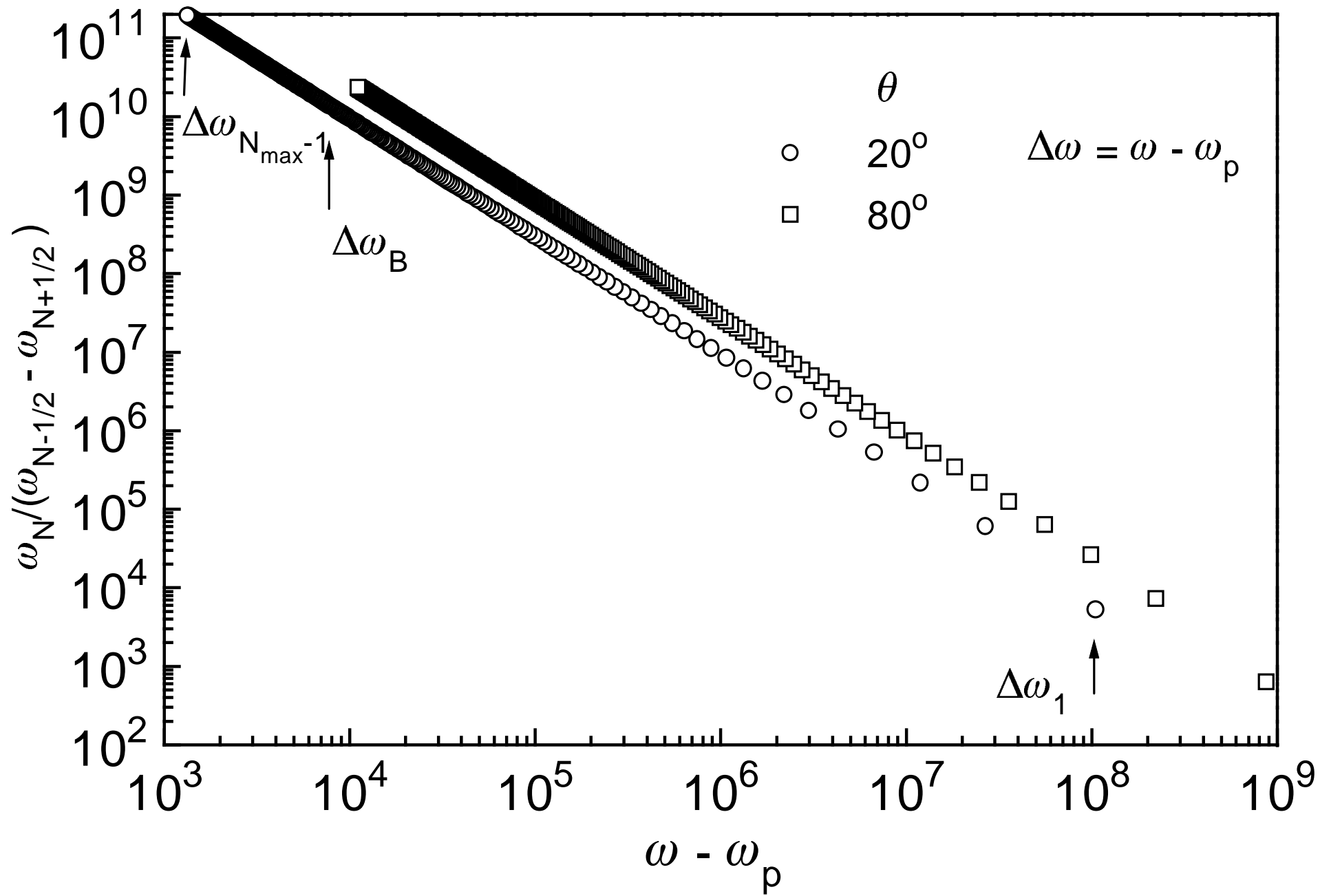
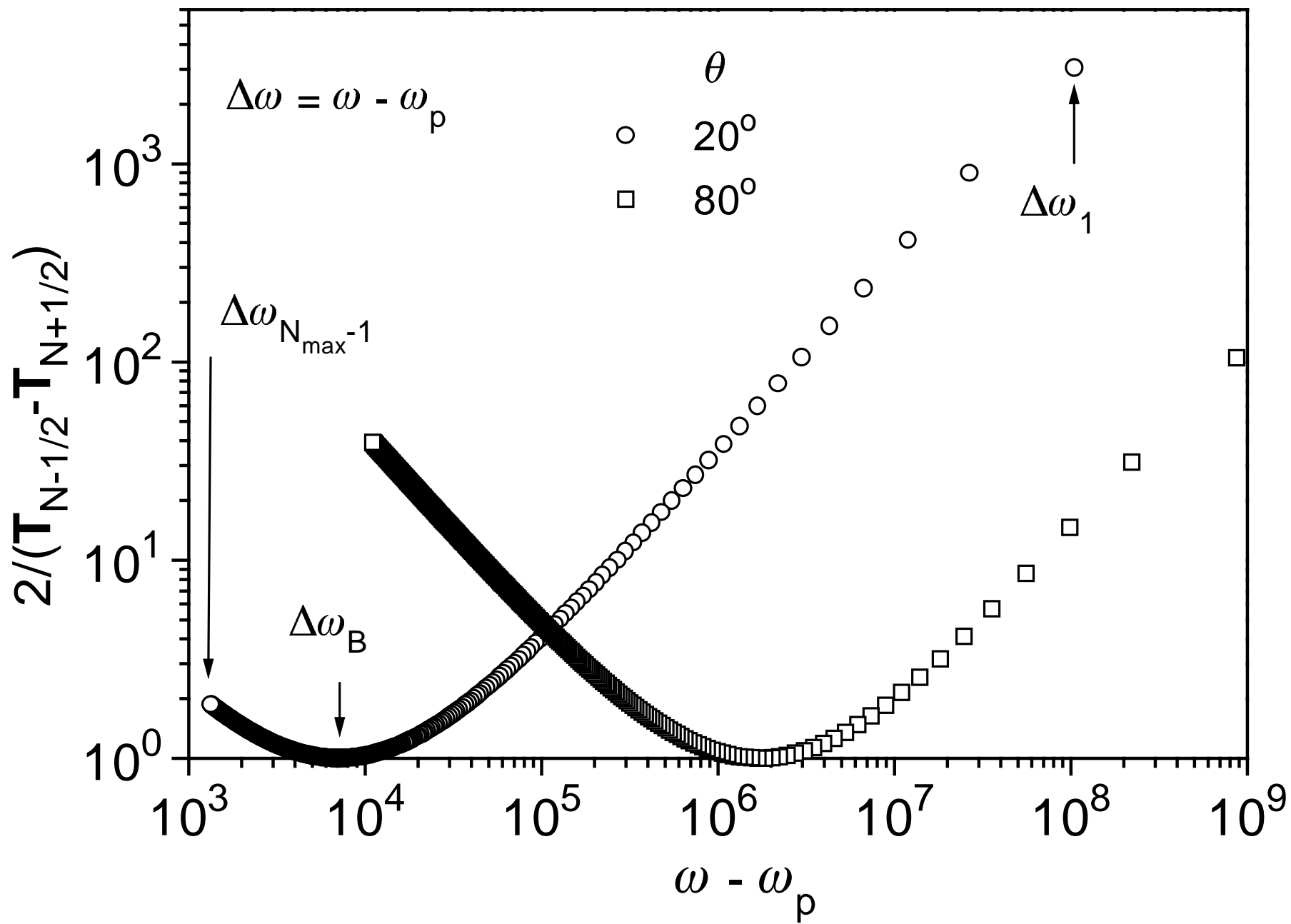
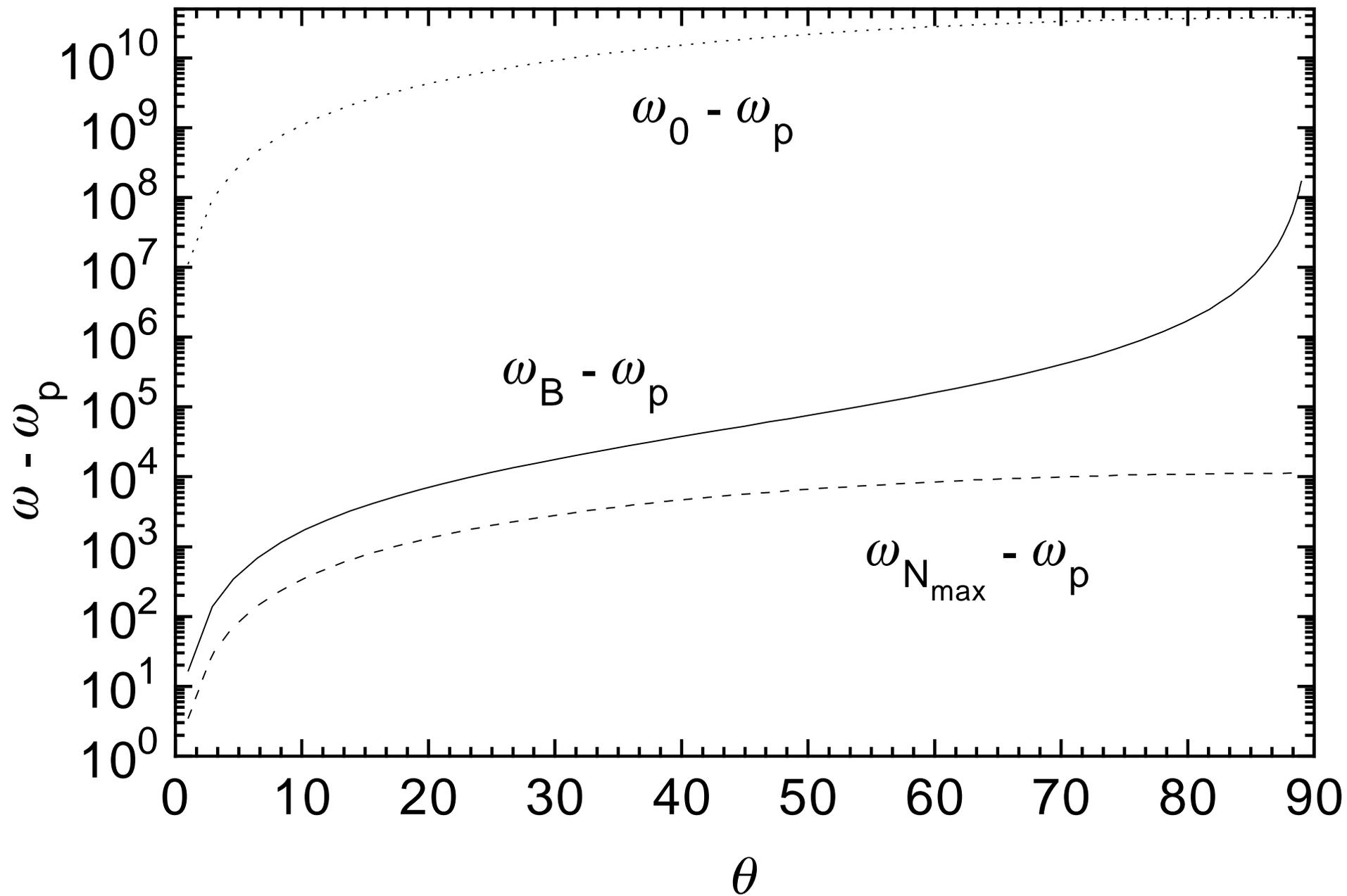
\end{document}